\documentclass[final]{usenixconf}
\usepackage[draft]{hyperref}
\usepackage[final]{textweaks}
\usepackage{comment}
\usepackage{upgreek}
\usepackage{url}
\usepackage{xspace}
\usepackage[small,bf]{caption}
\usepackage{listings}
\usepackage{graphicx}
\usepackage{enumerate}
\usepackage{booktabs}
\usepackage[compact]{titlesec}
\usepackage{algorithm}
\usepackage{algpseudocode}
\usepackage{amsmath}
\usepackage[svgnames,table]{xcolor}
\usepackage{tikz}
\usepackage{pgfplots}
\usepackage{authblk}
\usepackage{tabularx}

\pgfplotsset{
    ylabel right/.style={
        after end axis/.append code={
            \node [rotate=90, anchor=north] at (rel axis cs:1,0.5) {#1};
        }   
    }
}

\usepackage{caption}
\usepackage{subcaption}

\newcommand{\StatexIndent}[1][3]{%
  \setlength\@tempdima{\algorithmicindent}%
  \Statex\hskip\dimexpr#1\@tempdima\relax}

\newcommand{\name}{Flashield\xspace}

\title{\name: a Key-value Cache that Minimizes Writes to Flash}

\usepackage{authblk}
\makeatletter
\renewcommand\AB@affilsepx{, \protect\Affilfont}
\makeatother

\author[1]{Assaf Eisenman}
\author[1]{Asaf Cidon}
\author[2]{Evgenya Pergament}
\author[1]{Or Haimovich}
\author[3]{Ryan Stutsman}
\author[4]{Mohammad Alizadeh}
\author[1]{Sachin Katti}
\affil[1]{Stanford University}
\affil[2]{Technion}
\affil[3]{University of Utah}
\affil[4]{MIT CSAIL}

\begin{document}

\maketitle{}

\section*{Abstract}

As its price per bit drops, SSD is increasingly becoming the default storage
medium for cloud application databases.
However, it has not become the preferred storage medium
for key-value caches, even though SSD offers more than 10$\times$ lower
price per bit and sufficient performance compared to DRAM.
This is because key-value caches need to frequently insert, update and evict small objects.
This causes excessive writes and erasures on flash storage, since flash only supports
writes and erasures of large chunks of data.
These excessive writes and erasures
significantly shorten the lifetime of flash, rendering it impractical to use for key-value caches.
We present \name, a hybrid key-value cache that uses DRAM as a ``filter'' to minimize writes to SSD. \name performs light-weight machine learning profiling to predict which objects are likely to be read frequently before getting updated; these objects, which are prime candidates to be stored on SSD, are written to SSD in large chunks sequentially.
In order to efficiently utilize the cache's available memory,
we design a novel in-memory
index for the variable-sized objects stored on flash that requires only 4~bytes
per object in DRAM.
We describe \name's design and implementation and, we evaluate it on a
real-world cache trace.  Compared to state-of-the-art systems that suffer a
write amplification of 2.5$\times$ or more, \name maintains a median write amplification
of 0.5$\times$ without any loss of hit rate or throughput.

\section{Introduction}

Flash has an order of magnitude lower cost per bit of storage compared to DRAM.
Consequently it has become the preferred storage medium for hot data that requires high throughput and low latency access.
For example Google~\cite{googleflash} and Facebook~\cite{facebookflash} use it for storing photos,
and databases like LevelDB~\cite{leveldb} and RocksDB~\cite{rocksdb}
are deployed on top of flash.

However, flash is not used for key-value caches, an essential infrastructure tier
for modern web scale applications.
This is surprising because these caches are
typically deployed in a dedicated remote cluster~\cite{nishtala2013scaling} or
on a physically remote data center~\cite{memcachier,elasticache}. As a result, all accesses incur a network
access time of 100~$\upmu$s or more, hence flash can provide essentially the same access
latency as DRAM. Furthermore, since the performance of caches is primarily determined by the amount of memory capacity
they provide~\cite{cliffhanger,dynacache}, and the
cost per bit of SSD is more than 10$\times$ lower than DRAM, flash
promises significant financial benefits compared to DRAM. Table~\ref{tbl:costs} demonstrates the cost difference between DRAM-only cache and hybrid cache, both with 4.25~TB capacity. The total TCO difference would be even
greater due to power costs, since flash consumes significantly lower power compared to DRAM.
%Asaf:
%This whole "For Example" part might be too detailed. Maybe we can replace it with a figure showing the original configuration (server + DRAM) and the new configuration (server + DRAM + flash) with a dollar figure below each one of them to explain the difference

\begin{table}[b!]
\centering
\footnotesize
\setlength\tabcolsep{3.5pt}
\begin{tabularx}{\columnwidth}{X|cc|cc}
\hline
& \multicolumn{2}{c|}{Hybrid cache} & \multicolumn{2}{c}{DRAM-only cache} \\
& Count & Cost & Count & Cost \\
\hline
%\multirow{2}{*}{150,264}& \multirow{2}{*}{270,437} & \multirow{2}{*}{239,191} & \multirow{2}{*}{275,379}\\
Dell 2$\times$10 core server with 256~GB DRAM & \multirow{2}{*}{1} &\multirow{2}{*}{\$7700}& \multirow{2}{*}{17} & \multirow{2}{*}{\$130,900} \\
\hline
Samsung 1~TB enterprise SSD & \multirow{2}{*}{4} & \multirow{2}{*}{\$4800} & \multirow{2}{*}{0} & \multirow{2}{*}{0}\\
\hline
Total & & \$12,500 &  & \$130,900 \\
\hline
\end{tabularx}
\caption{The cost of a hybrid SSD and DRAM cache server with combined capacity of 4.25~TB, versus the cost of multiple DRAM-only cache servers with the same aggregate capacity.}
\label{tbl:costs}
\end{table}

%For example, a Dell 2$\times$10 core server with 256~GB of DRAM costs about \$7,700, and a
%Samsung 1~TB enterprise drive costs about \$1,200. Therefore, the combined DRAM and SSD capacity of a server with four
%attached drives would be 4.25~TB at a list price of \$12,500. To match that capacity with
%the aggregate DRAM of multiple servers would cost \$130,900. The total TCO %difference would be even
%greater due to power costs, since flash consumes significantly lower power compared to DRAM.

The reason flash has not been adopted as a key-value cache is that cache workloads wear out flash drives very quickly. These workloads typically
consist of small objects, some of which need to be frequently updated~\cite{nishtala2013scaling,atikoglu2012workload}.
But flash chips within SSDs can only be written a few thousand times per
location over their lifetime. Further, SSDs suffer from
write amplification (WA).
%Asaf:
%I don't think the DLWA explanation here is clear. Maybe we can just say that flash needs to be written to in large chunks to avoid DLWA, because...
That is, for each cache object write, several more bytes are written to the actual flash chips at the device level. The reason is that flash pages are physically grouped in large blocks. Pages must be erased before they can be overwritten, but that can only be done in the granularity of blocks. The result is that over time, these large blocks typically contain a mix of valid pages and pages whose contents have been invalidated.
Any valid pages must be copied to other flash
blocks before a block can be erased. This garbage collection process creates
device-level write amplification (DLWA) that can increase the amount of data
written to flash by orders of magnitude. Modern SSDs exacerbate
this by striping many flash blocks together (512~MB worth or more) to increase
sequential write performance (\S\ref{sec:device-wa},~\cite{RIPQ}).

To minimize the number of flash writes, SSD storage systems are constrained to writing data in
large contiguous chunks.  This forces a second-order form of write
amplification called {\em cache level write amplification} (CLWA).  CLWA occurs
when the cache is forced to relocate objects to avoid
DLWA.  For example, when a hot object occupies the same flash block as many
items that are ready for eviction, the cache faces a choice. It can evict the
hot object with the cold objects, or it can rewrite the hot object as part of a
new, large write. Therefore, in existing SSD cache designs, objects
get re-written multiple times into flash.
These challenges will become even greater over time, since it is projected that as
flash density increases, its durability will continue to decrease~\cite{bleak}.
% Data center operators typically provision their hardware for a certain number of years
% (e.g., three years~\cite{googleflash,facebookflash}), and the unpredictable lifetime of running a cache on SSD
% incurs increased hardware
% cost and makes SSDs difficult to maintain by data center operators.
% Therefore, SSD-based key-value caches that suffer from a high rate of WA, will
% become even more difficult to adopt.

We present \name, a design for a hybrid key-value cache that uses both DRAM and SSDs.
Our contribution is a novel caching strategy that significantly extends the lifetime of SSDs
such that it is comparable to DRAM by minimizing the number of writes to flash.
Our main observation is that not all objects
entering the cache are good candidates for placement in SSD. In particular, the cache should avoid writing
objects to flash that will be updated or that will not be read in the near future.
However, when objects first enter
the cache, it does not know which objects are good candidates for SSD and which are not.
Therefore, the key idea in \name's design is that incoming objects into the cache
always spend a period of time in DRAM, during which the cache learns whether they are good candidates for
flash storage. If they indeed prove themselves as flash-worthy,
\name will move them into flash. If not, they are never moved into flash, which minimizes
the resulting write amplification.
Since the flash layer is considerably
larger than DRAM (e.g., 10$\times$ larger), objects moved to flash on average will remain in the
cache much longer than those that stay in DRAM.

To dynamically decide which objects are flash-worthy under varying workloads,
we implement the filtering algorithm using machine-learning based
Support Vector Machine (SVM) classification. We train a different classifier for each application in the cache.
To train the classifiers, we design a light-weight sampling
technique that uniformly samples objects over time, collecting statistics about the number of past accesses
and the time between accesses. The classifier is used to predict whether an object
will be read more than $t$ times in the future and
determine its suitability to be stored in flash. We term this metric \emph{flashiness}.

The second main idea in \name's design is its novel DRAM-based lookup index for variable-length objects
stored on flash that only requires less than 4 bytes of DRAM per object.
Since the flash layer's capacity is much larger than the DRAM's, a na\"{\i}ve lookup index for objects
stored on flash
would consume the entire capacity of the DRAM. Our index consumes a relatively small amount of memory
by not storing the location of the objects and their corresponding keys.
Instead, for each object stored on flash, the index contains a pointer to a region in the flash where the object is stored,
and it stores an additional 4~bits that specify a hash function on the object key that indicates the insertion point of the
object in its region on flash.
The index leverages bloom filters to indicate whether the object resides on flash or not without storing full keys in DRAM.
On average, \name's lookup index only requires 1.03~reads from the SSD to return an object stored on it.

We implement \name in C and evaluate its performance under a commercial trace. We show that compared with RIPQ~\cite{RIPQ},
the state-of-the-art SSD key-value cache, \name reduces write amplification by a median of 5$\times$ and an average of 16$\times$, while
maintaining the same average hit rates.
We show that when objects are read from SSD, \name's read latency and throughput is
close to the SSD's latency and throughput, and when objects are written to the cache or read from DRAM,
its latency and throughput are similar to that of DRAM-based caches.

This paper makes three main contributions:
\begin{enumerate}
\itemsep0em
\item \name is the first SSD storage system which uses DRAM as a filter for deciding which objects
to insert into flash.
\item \name's novel flash lookup index takes up less than 4 bytes per object in DRAM by not storing the direct location of objects and their keys.
\item \name is the first key-value cache that uses a machine-learning based algorithm and lightweight temporal sampling to predict which objects will be good candidates for flash.
\end{enumerate}

As new generations of flash technology can
tolerate even fewer writes~\cite{bleak}, our dynamic admission control to flash can be extended
to other systems beyond caches, such as flash databases and file systems.

\section{The Problem}

Building a cost-effective SSD-based cache requires solving two conflicting challenges. SSDs perform poorly and wear out quickly unless writes
are large and sequential. The lifetime of an SSD is defined by flash device manufacturers as the amount
of time before a device has a non-negligible probability of producing
uncorrectable read errors. The lifetime of an SSD depends on several factors, including the number of writes and erasures (termed program-erase
cycles), the average time between refresh cycles of the SSD cells, the cell
technology, the error correction code and more. The typical lifetime of a flash cell is between 3-5 years assuming it is written 3-5 times a day on average.
This conflicts with the characteristics of cache workloads. Caches store small objects with highly variable lifetimes; this drives caches to prefer small random I/O for reads and writes which will wear flash drives out quickly.

\begin{figure}[t]
    \centering
	\includegraphics[width=\columnwidth]{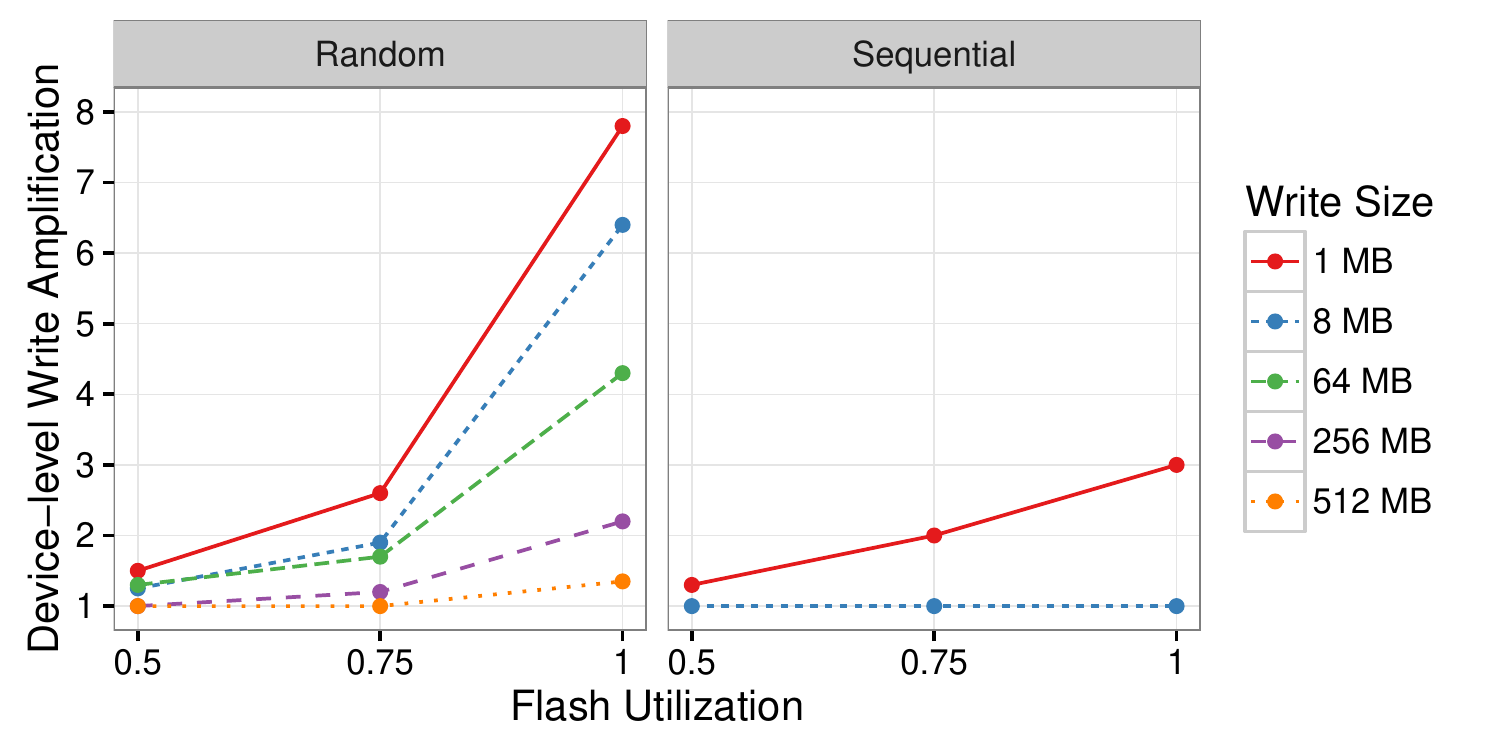}
    \caption{Device-level write amplification after writing 4~TB randomly and sequentially.}%
    \label{fig:device-level}%
\end{figure}

The key metric that helps us track device wear is write amplification. Many write patterns force the SSD to perform
additional writes to flash in order to reorganize data.  The ratio of the
bytes written to flash chips compared to the bytes sent to the SSD by the
application is called \emph{write amplification} (WA). A WA of 1.0 means each
byte written by the application caused a one byte write to flash. A WA of 10.0
means each byte written by the application caused an extra 9~bytes of data to be
reorganized and rewritten to flash.

\subsection{Device-level Write Amplification}
\label{sec:device-wa}

\emph{Device-level write amplification} (DLWA) is write amplification that is caused
by the internal reorganization of the SSD.
The main source of DLWA comes from the size of the unit of flash
reuse. Flash is read and written in small (\textasciitilde{}8~KB) pages.
However, pages cannot be rewritten without first being erased. Erasure happens
at a granularity of groups of several pages called blocks
(\textasciitilde{}256~KB). The mismatch between the page size (or object sizes)
and the erase unit size induces write amplification when the device is at high
utilization.

For example, when an application overwrites the contents of a page, the SSD
writes it to a different, fresh block and maintains a relocation mapping called
the flash translation layer (FTL). The original block cannot be erased yet,
because the other pages in the same block may still be live.  When the flash
chips are completely occupied, the SSD must erase blocks in order to make room
for newly written pages. If there are no blocks where all of the pages have
been superseded by more recently written data, then live pages from several blocks
must be consolidated into a single flash block.
This consolidation or {\em garbage collection} is the source of DLWA. If a device is at 90\% occupancy, then its DLWA
can be very high. Figure~\ref{fig:device-level} measures this effect. It shows DLWA under sequential and random writes. The measurements were
taken on a 480~GB Intel~535 Series SSD using SMART.  For each data
point, 4~TB of randomly generated data is written either randomly or
sequentially to the raw logical block addresses of the device with varying
buffer sizes. Specifically, in the random workload the logical block space
is broken into contiguous fixed buffer-sized regions; each write overwrites one of the
regions at random with a full buffer of random data. The sequential workload is
circular; regions are overwritten in order of their logical block addresses,
looping back to the start of the device as needed.  For both patterns, we
varied the space utilization of the device by limiting writes to a smaller
portion of the device's logical block addresses.

The results show that random, aligned 1~MB flash writes experience a nearly
8$\times$ DLWA. This is surprising, since flash
erase blocks are smaller than 1~MB. The reason for this write amplification is
because SSDs are increasingly optimized for high write bandwidth. Each flash
package within an SSD is accessed via a relatively slow link (50-90~MB/s today);
SSDs stripe large sequential writes across many flash packages in parallel to
get high write bandwidth. This effectively fuses several erase blocks from
several packages into one logical erase block. A 1~MB random write marks a
large region of pages as ready for erase, but that region is striped across
several erase units that still contain mostly live pages.
Others have corroborated this effect as well~\cite{RIPQ}.

There are two ways to combat this effect. The first is to write in units of $B
\cdot{} W$ where $B$ is the erase block size and $W$ is how many blocks the SSD
stripes writes across. Our results show that a cache would have to write in
blocks of 512~MB in order to eliminate DLWA. The
second approach is to write the device sequentially, in FIFO-order at all
times. This works because each $B \cdot{} W$ written produces one completely
empty $B \cdot{} W$ unit, even if writes are issued in units smaller than $B
\cdot{} W$. Figure~\ref{fig:device-level} shows that 8~MB sequential writes
also eliminate DLWA.

This means our cache is extremely constrained in how it writes data
to flash. To minimize DLWA the cache must write
objects in large blocks or sequentially. In either case, this gives the cache
little control on precisely {\em which} objects should be replaced on flash.

\subsection{Cache-level Write Amplification}
\label{sec:cache-wa}

Writing to flash in large \emph{segments} (contiguous chunks of data) is a
necessary but not sufficient condition for minimizing the overall SSD write amplification.
The main side effect of writing in large segments is \emph{cache-level write amplification} (CLWA).
CLWA occurs when objects that were removed
from the SSD are re-written to it by the cache eviction policy.
If the size of the segments (MBs) is significantly larger than the size
of objects (bytes or KBs),
it is difficult to guarantee that high-ranking objects in the cache will
always be stored physically separate from low-ranked objects or objects that contain old values.
Therefore, when a segment that has
many low-ranked objects is erased from the cache, it may also
inadvertently erase some high-ranking objects.

\begin{table}[t!]
\centering
\footnotesize
\begin{tabular}{rrr}
\toprule
Avg Object Size & Read/Write/Update \% & Unread Writes \%\\  \midrule
257~B & 90.0\%/9.5\%/0.5\% & 60.6\%\\
\bottomrule
\end{tabular}
\caption{Statistics of the top 20 most applications with the most requests in the week-long Memcachier trace. 90\% of all requests are reads, 9.5\% are writes and 0.5\% are updates. 60.6\% of writes are never read after they are written.}
\label{tbl:stats}
\end{table}

\begin{figure}[t]
    \centering
	\includegraphics[width=\columnwidth]{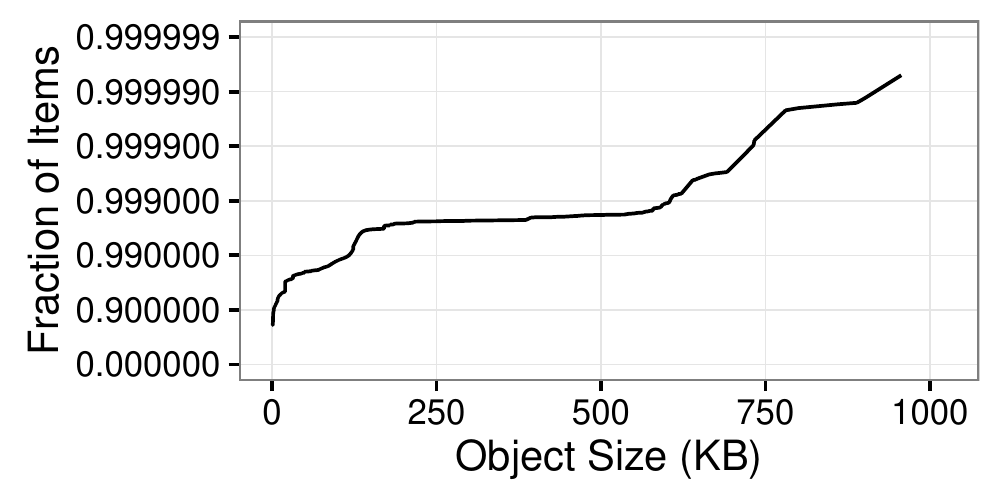}
    \caption{CDF of the object sizes written to memory by the top 20 applications in the Memcachier trace.}%
    \label{fig:cdf}%
\end{figure}

Table~\ref{tbl:stats} presents general statistics of a week-long trace of Memcachier,
a commercial Memcached service provider~\cite{dynacache,cliffhanger}, and
Figure~\ref{fig:cdf} presents the distribution of the sizes of objects written in the trace.
The figure demonstrates that object sizes vary widely, and in general they are
very small: the average size of objects written to the cache is 257 bytes, and 80.67\% of
objects are smaller than 1~KB.
Therefore, even with a segment size of 8~MB using sequential writes, which is the
the smallest possible segment size that does not incur extra write amplification,
each segment will contain on average over 32,000 unique objects.

In addition, 60.6\% of writes (and 5.8\% of all requests) are unread writes, which means they are never read after they are written,
and 0.5\% of all requests are updates.
Both unread writes and updates contribute to write amplification.
Ideally, objects that will never be read should not be written to the cache.
In the case of updates, to reclaim the space of an object after it was updated,
the cache needs to erase and write the value of the object.

The state-of-the-art system, RIPQ~\cite{RIPQ}, an SSD-based photo caching system that caches
large immutable objects, tries to minimize CLWA by
inserting objects that were read $k$ times in the past together.
When objects are first inserted into the cache, they are buffered in memory, and periodically
they are moved into flash together as a segment with other objects that have been read the same number of times.
The idea is that objects that were read $k$ times in the past might share
a similar future eviction rank.
For example, an object that was read once is stored on flash in the same segment
with other objects that were read once. The rationale is that objects in the same segment will have a similar lifetime,
so when it comes time to evict the segment, most of its objects will be cold and will not have to be re-written.
In addition, segments that contain objects that have been read fewer times will be evicted faster than
segments with objects that have been read many times.

\begin{table}[t!]
\centering
\footnotesize
\begin{tabular}{rrr}
\toprule
& Hit Rate & CLWA \\  \midrule
Victim Cache & 69.72\% & 4.00 \\
RIPQ & 70.59\% & 2.59 \\
\bottomrule
\end{tabular}
\caption{Hit rate and cache-level write amplification of RIPQ and the victim cache policy under the entire Memcachier trace.}
\label{tbl:RIPQ}
\end{table}

In order to test this strategy, we simulated the CLWA of RIPQ (the implementation is not publicly available) with
the Memcachier trace, using a segmented LRU with 8 queues. We also compared it with a victim cache policy,
a na\"{\i}ve approach
where the SSD simply serves as an L2 cache (i.e., every object evicted from DRAM is written to SSD).
This policy is used by TAO~\cite{tao}, Facebook's graph data store, which leverages a limited amount of flash as a victim cache for data stored in DRAM.
The simulation assigns the same amount of memory for each application in the trace, with a ratio of DRAM to SSD of 1:7.
So for example, if an application was originally assigned 1~GB in the trace, the simulation would assign it a capacity of
128~MB of DRAM and 896~MB of SSD.

The results of the simulation are presented in Table~\ref{tbl:RIPQ}. The results show that while RIPQ considerably
improves upon victim cache, it still suffers from a very high CLWA. Note that the victim cache would suffer
from an even greater total WA, because it also suffers from DLWA (since it does not write to flash in large segments).
The reason RIPQ suffers from CLWA, is twofold.
First and most importantly, RIPQ automatically writes all incoming objects to flash. Even objects that will never be read again or are
frequently updated, will be written to flash.
Second, when the frequency of reads of a certain object changes, it creates additional writes.
For example, if an object was read twice over a period of time after it was written, it is grouped with other objects that were read twice on flash. However, if a burst occurred and it was read five more times, RIPQ needs to rewrite it to group it with other higher ranking objects.
Since the objects are much smaller than the segment size, and there is a relatively high ratio of writes in the trace,
RIPQ struggles to guarantee that objects that have been read around the same time will be stored in the same segment.

This example teaches us two lessons on how to minimize CLWA.
First, not every object that is written by the application to the cache, should necessarily be stored on SSD. For example, objects
that are updated soon after they are first written or objects that have a low likelihood of being read in the future.
However, the occurrence of such objects varies widely across different applications. For example, in some applications of the Memcachier trace,
more than half of written objects
are never read again, and in some applications, a vast majority of objects are read many times and should be written to the cache.
Second, due to the disparity between the segment size and the object size,
it is difficult to guarantee that objects that were similarly ranked by the eviction policy will be stored in physically adjacent regions on SSD.

Both of these insights motivate us to design \name, a cache that successfully minimizes CLWA.

\section{Design}

\begin{figure}[t]
    \centering
	\includegraphics[width=\columnwidth]{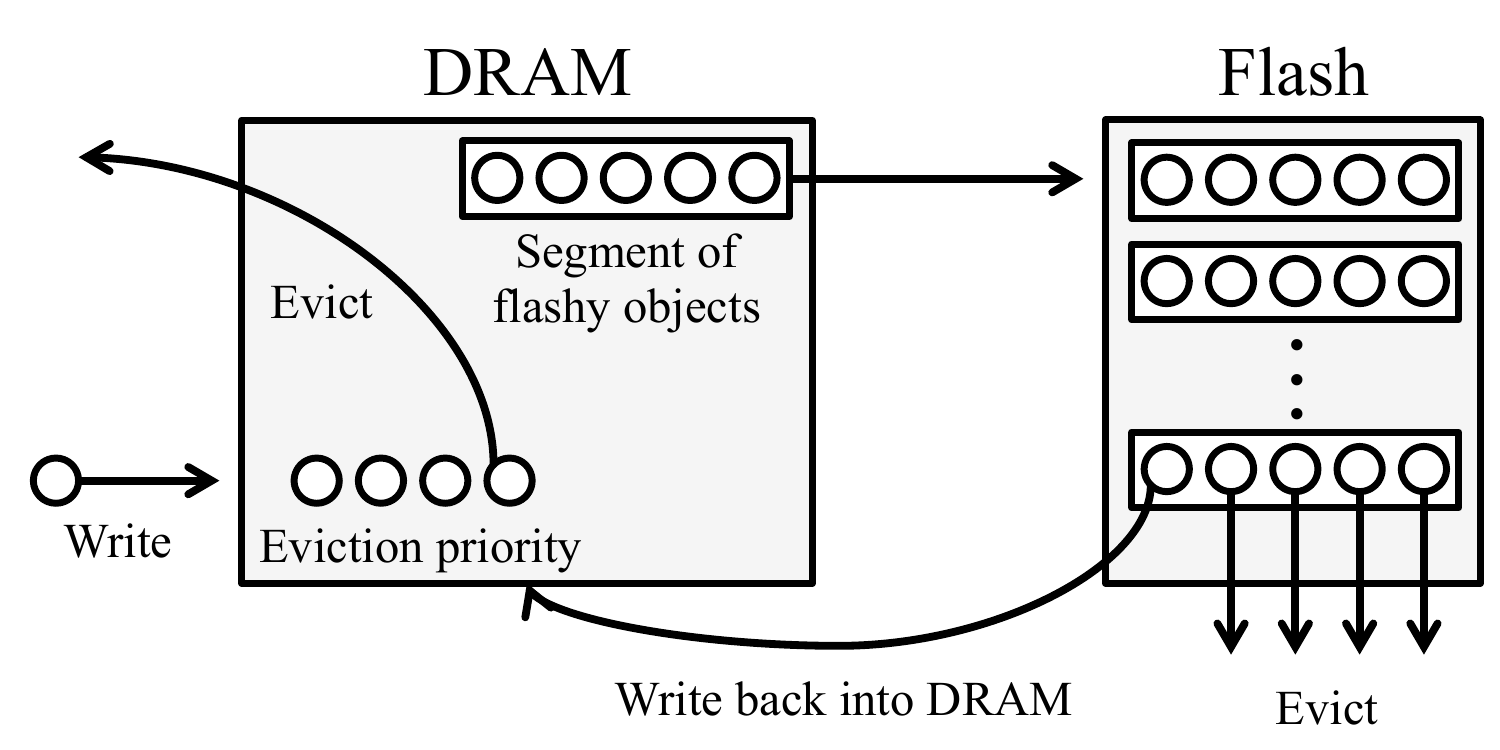}
    \caption{Lifetime of an object in \name. Objects always enter into DRAM. Objects that are a good fit for flash (\emph{flashy} objects) are
    aggregated and moved into flash as a segment. The decision of whether to evict objects from DRAM or flash is based on a global
    eviction priority.}%
    \label{fig:lifetime}%
\end{figure}

This section presents \name's design.
The design goal is to minimize cache-level write amplification, while maintaining
the highest possible hit rate. 
\name is a hybrid key-value cache that uses both SSD and DRAM.
The key insight of \name's design is to use DRAM as a filter, which prevents
moving objects into flash that will be soon thereafter evicted or updated.

Figure~\ref{fig:lifetime} illustrates the lifetime of an object in \name. Objects are first always written to DRAM.
After the object is read for the first time, \name starts collecting features that describe its performance.
These contain information about how many times and how frequently the object has been accessed.
At any point in time during its duration in DRAM, an object may be evicted by \name's eviction
algorithm.
Periodically, \name moves a segment (e.g. 512~MB) composed of many DRAM objects into flash. \name utilizes a machine learning
classifier to rank the objects based on their features. If the object passes a rank threshold,
it will be considered as a candidate to move to flash. The candidates to flash are then ranked
based on their score, which determines the order they are moved by \name into flash.
After it gets moved to flash, an object will live in the cache for a relatively long duration.
It will get moved out of flash once its segment is erased from flash, in FIFO order.
At that point, the object will be evicted if it is low in terms of eviction priority, or it will
get re-inserted into DRAM if it has a high eviction priority.

In \name, DRAM serves three purposes. First, it is used as a filter to decide which objects
should be inserted into SSD. Second, it stores the metadata for looking up objects on flash.
Third, it serves as a caching layer for objects before they are moved to SSD
and for objects that are not candidates for SSD. 
In the rest of this section, we focus on the first and second roles of DRAM. 

\subsection{DRAM as a Filter}

In \name, DRAM serves as a proving ground for moving objects into flash.
When objects are first written into DRAM, \name does not have any a-priori knowledge whether they
will be good candidates for flash. Furthermore, given the great diversity of applications that
utilize key-value caches, applications have varying access patterns.

A strawman approach for determining which objects are flash-worthy is to rank them based on
simple metrics like time-to-last-access or access frequency, as done by standard cache replacement 
polices like LRU or LFU. However, simply ranking objects is insufficient, because it is difficult to
set a single threshold for flash-worthiness that will work for all applications.
For example, we can set a threshold requiring that an object will be read more than once
before it enters flash. Such a threshold proves too stringent for certain applications where the access
patterns are long and may cause excessive misses
due to premature evictions, and too lenient for other applications where many objects would be unnecessarily written to flash.

Instead of using a one-size-fits-all approach, machine learning can be used as a way to
dynamically learn which objects are a good
fit for flash for each individual application.
In order to apply a machine learning classifier, we need to define the metric
we are trying to estimate and the features
that can predict the metric.

We define \emph{flashiness} as a metric that predicts whether an object will be a good fit for flash.
An object that has a high flashiness score is an object that meets two criteria. First, it is an object
that will be accessed several times in the near future. This guarantees that it will not be
evicted by the cache's eviction function. Second, it needs to be immutable in the near future,
since updating an object in SSD requires an additional write and erasure.

Both of these criteria can be captured by predicting the number of times an object will be read
in the near future (e.g., one hour), while it is stored on DRAM. If the object is evicted or updated
during this period, we only count the number of reads until the object was evicted or updated.
Initially, we tried predicting this number using a logistic regression.
We ran this classifier on a commercial Memcachier trace and found the prediction was highly inaccurate.
After trying different features and classifiers,  we found it is difficult to accurately predict how many times an
object will be accessed in the future.

\begin{table}[t!]
\centering
\footnotesize
\begin{tabular}{rrr}
\toprule
App & Accuracy & Recall \\ \midrule
1 & 39.0\% & 100.0\% \\
3 & 92.2\% & 100.0\%\\
19 & 95.3\% & 100.0\%\\
18 & 96.9\% & 100.0\%\\
20 & 77.6\% & 100.0\%\\
\bottomrule
\end{tabular}
\caption{Accuracy and recall of SVM classifier for predicting if an object will be accessed at least once in the future, for the top 5 applications in the Memcachier trace in terms of number of requests.}
\label{tbl:accuracy}
\end{table}

Therefore, instead of predicting the number of times an object will be accessed in the near future,
\name uses a binary classifier, using Support Vector Machine (SVM), which predicts
whether an object will be accessed more than $n$ times in the near future.
Table~\ref{tbl:accuracy} provides the accuracy ($\dfrac{tp}{tp+fp}$, where $tp$ is true positives and $fp$ is false positives)
and recall ($\dfrac{tp}{tp+fn}$, where $fn$ is false negatives) for the classifier when it tries to predict whether
an object will be accessed at least once in the future, using a training time of one day.
Note that the accuracy varies widely across applications when the recall is 100\%. This indicates that
for certain applications (e.g., application 1), it is harder to accurately measure flashiness based on the history
of requests than for other applications.

We experimented with several different features related to the number and frequency of object requests. 
Based on that, our design uses the following five features: number of past reads to the object,
the average time between these reads,
the time between the last two reads, the maximum time between subsequent reads and the time it took for the first
read after the object
was written.

\begin{figure*}[t!]
\begin{subfigure}{.333\linewidth}
\centering
\resizebox{1.136\linewidth}{!}{\includegraphics{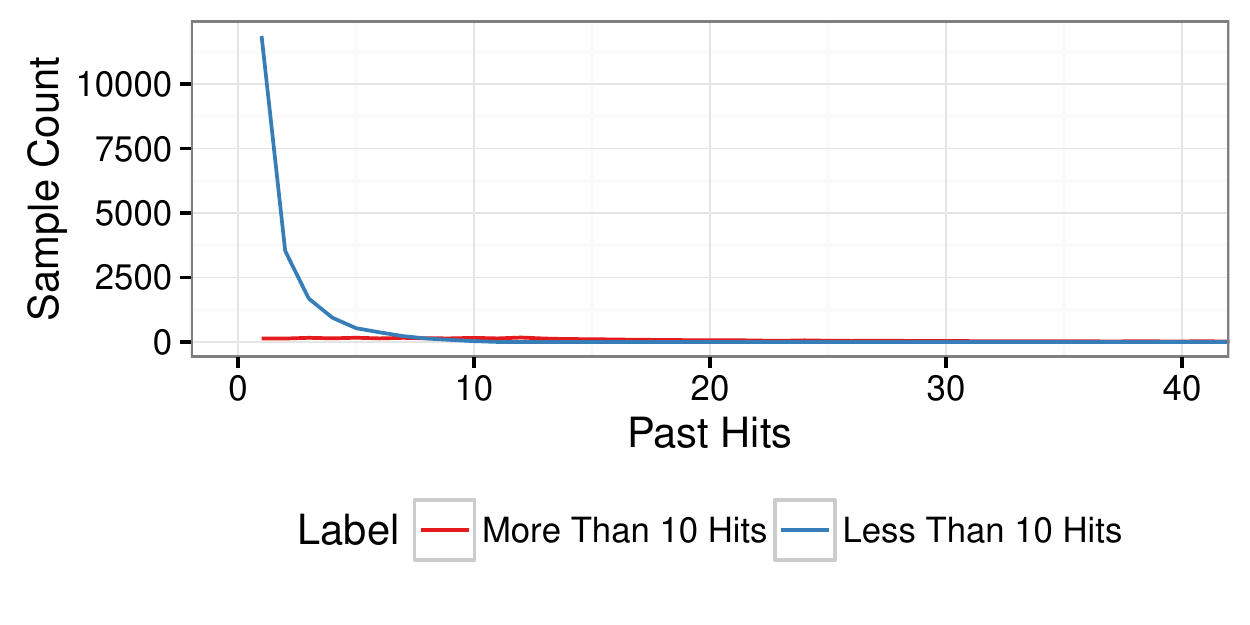}}
\caption{}
\label{fig:sub1}
\end{subfigure}%
\begin{subfigure}{.333\linewidth}
\centering
\resizebox{1.136\linewidth}{!}{\includegraphics{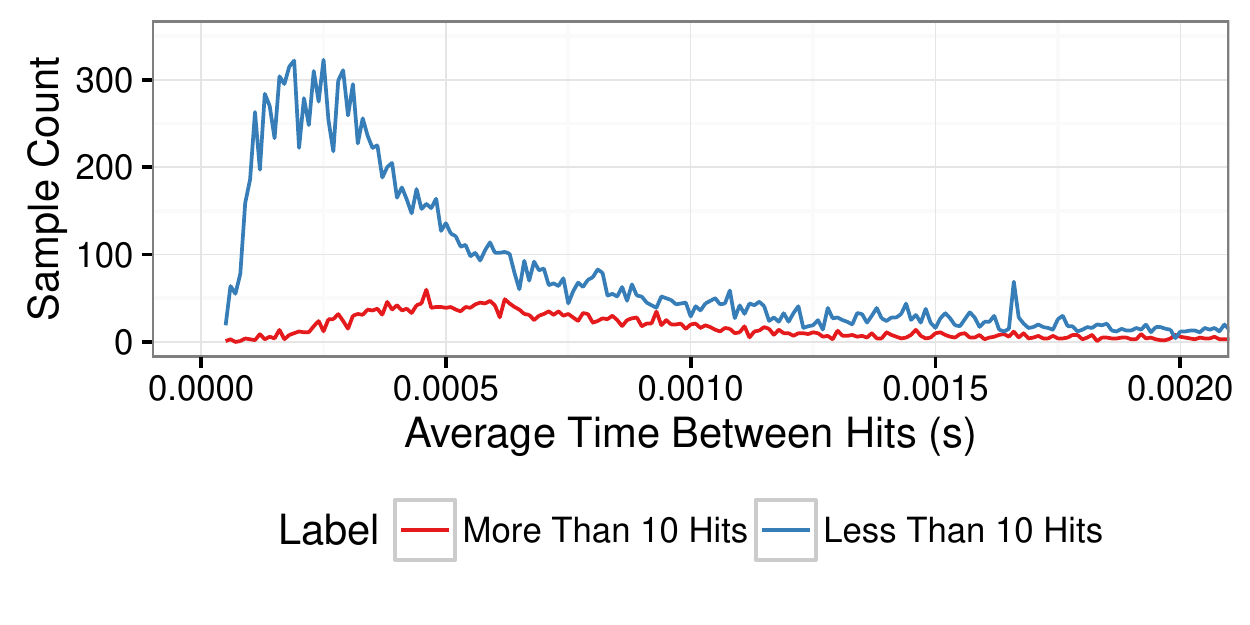}}
\caption{}
\label{fig:sub1}
\end{subfigure}%
\begin{subfigure}{.333\linewidth}
\centering
\resizebox{1.136\linewidth}{!}{\includegraphics{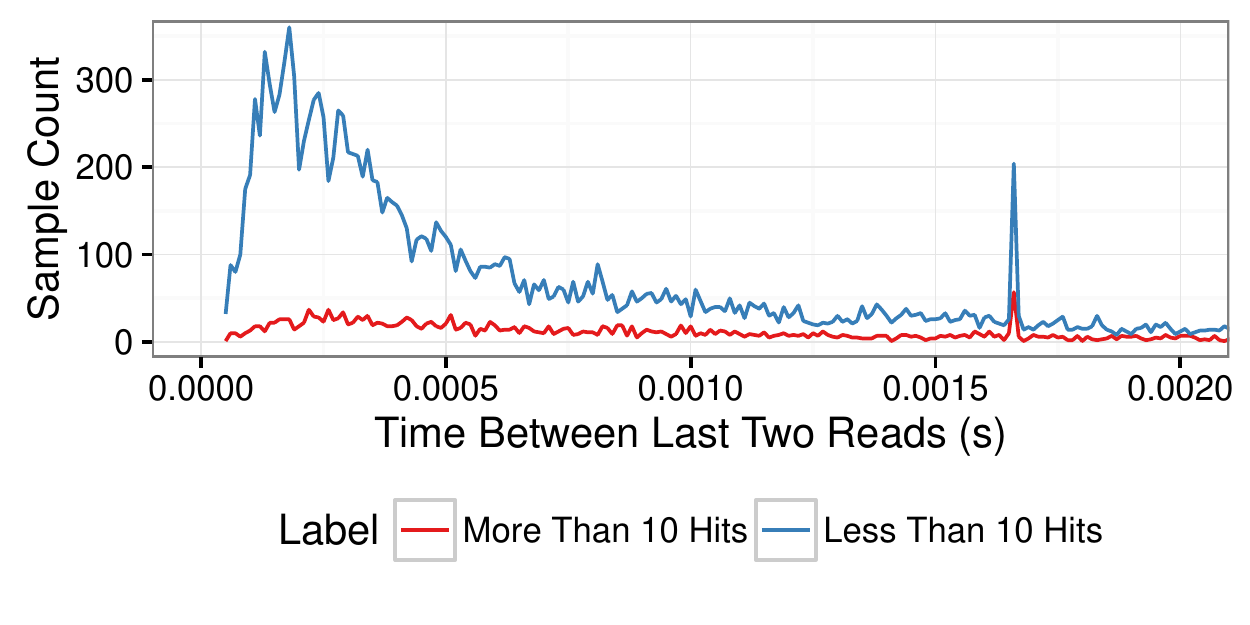}}
\caption{}
\label{fig:sub2}
\end{subfigure}\\[1ex]
\begin{subfigure}{.5\linewidth}
\centering
\resizebox{.75\linewidth}{!}{\includegraphics{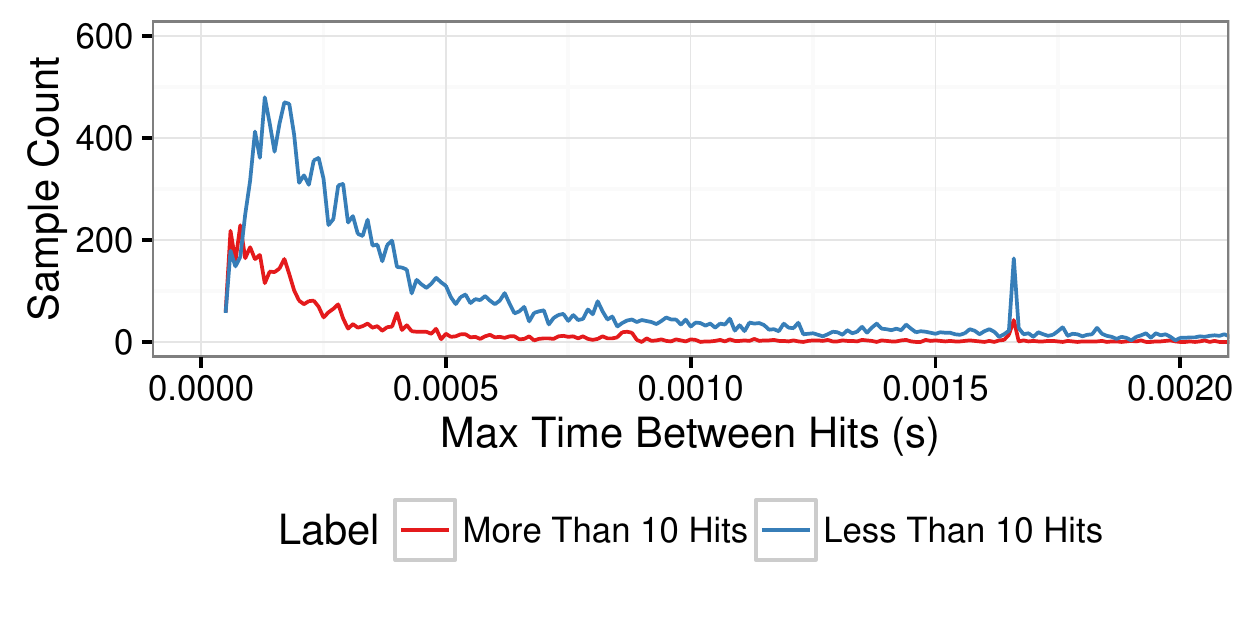}}
\caption{}
\label{fig:sub2}
\end{subfigure}
\begin{subfigure}{.5\linewidth}
\centering
\resizebox{.75\linewidth}{!}{\includegraphics{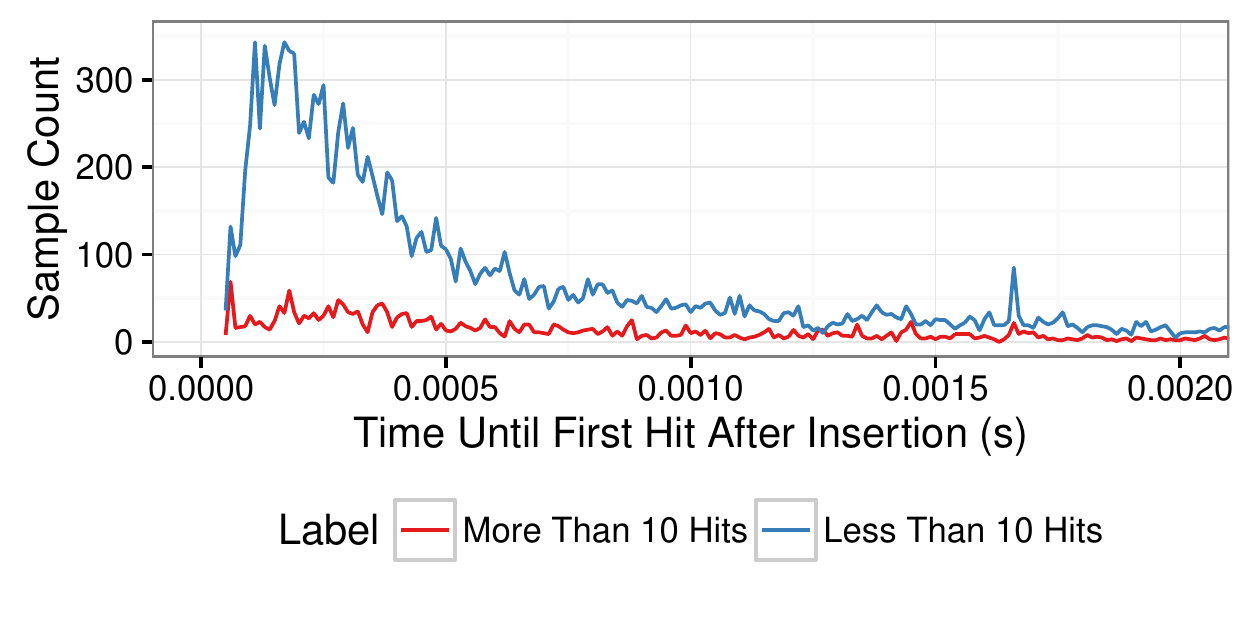}}
\caption{}
\label{fig:sub3}
\end{subfigure}
\caption{Prediction of whether an object will be read more than 10 times in the future with each of the five features, running application 19 from the Memcachier trace. Feature a is the number of past reads, feature b is the average time between reads, feature c is the time between
the last two reads, feature d is the maximum time between subsequent reads and feature e is the
time it took for the first read after the object was inserted.}
\label{fig:predictions}
\end{figure*}

%\begin{table}[t]
%\centering
%\footnotesize
%\begin{tabular}{rrrrrr}
%\toprule
%& \multicolumn{5}{c}{Features} \\
%App & a & b & c & d & e \\ \midrule
%1 & 1.7E+00 & -1.4E-04 & 9.3E-05 & 4.5E-04 & -4.1E-04 \\
%3 & 5.0E-01 & 4.5E-06 & 4.5E-06 & 1.1E-05 & -7.3E-06 \\
%19 & 3.8E-01	& 4.4E-05 & 4.4E-05 & 7.2E-05 & -6.7E-05 \\
%18 & 4.6E-01	& -1.1E-05 & -1.1E-05 & 2.6E-04 & -2.7E-04 \\
%20 & 4.0E-01 & 2.9E-05 & 2.9E-05 & 2.5E-04 & -3.4E-04 \\
%\bottomrule
%\end{tabular}
%\caption{Support Vector Machine co-efficients of the 5 features used on top 5 applications of the Memcachier trace.}
%\label{tbl:coefficients}
%\end{table}

Figure~\ref{fig:predictions} depicts the relationship of these features with predicting whether an object
will be accessed more than once in the next hour.  It buckets the number of future hits in the Y axis, as a function
of each feature on the X axis. 

Note that the threshold $n$, the number of times an object will be read in the future, can be used
by the system to indicate how sensitive it is to write amplification. If the system is very
sensitive to write amplification, it can set $n$ to a relatively high number (e.g., 10 or 100), which
will ensure that \name will only move objects into flash which it predicts will be read many times in the future.
On the other hand, if the system is more sensitive to hit rate, $n$ will be set as a low number (e.g., 1). 
In addition, \name allows the operator to set a fixed limit on the flash write rate to maintain a certain target lifetime (e.g., 5 years).

\subsection{DRAM as an Index for Flash}

The lookup indexes for flash and DRAM are both stored in memory. Since we use DRAM also as a filter,
the design goal of the indexes is that they will consume a minimal amount of space on DRAM.

We decided to use an in-memory index for objects stored in flash for two reasons.
First, since the index needs to be frequently updated, if it were stored in flash it would create significant write amplification.
Second, storing the index in DRAM more than halves the latency, since otherwise, each lookup for a key
would require an extra read from flash.

A na\"{\i}ve index would contain the identity of the keys stored in flash, the location
of the values, and their position in an eviction queue.
However, such an index would be prohibitively expensive. If we take an example of a 6~TB
flash device with an average object size of 257 bytes (equal to the average object size of the top 20 applications in the Memcachier trace), storing a hash of the key for each object that avoids collisions requires at least 8 bytes,
storing the exact location of each object would be 43 bits, and keeping a pointer to a position in a queue would be 4-8 bytes.
Storing 17 bytes per object on DRAM would require
406~GB of DRAM. This would take up (or exceed) all of the DRAM of a high end server.

Instead, we design a novel in-memory lookup index for variable-sized objects with an overhead of less than 4 bytes per object.
Rather than directly storing the location of the SSD object, the index has two separate fields: segment number and predefined hash function ID. The segment number points to a contiguous segment in flash where the object is stored. 
The output of the predefined hash function indicates the object location inside the segment.
%We elaborate on this technique in \S\ref{sec:writing}.
We chose to utilize 16 pre-defined hash functions since increasing the number of hash functions beyond that provided negligible improvement in the flash utilization. We explore the flash utilization in \S\ref{sec:utilization}.
Note that since data is written to flash sequentially, a segment sizes of 8~MB or larger achieves minimal DLWA. We use 512~MB segments in order to reduce the indexing overhead.

\name does not store the identities of keys in the index but instead only stores them in the flash device, as part of the object. In order to identify hash collisions in the lookup table, \name compares the key from flash. 
To limit the number of flash reads during key lookup and avoid complex table expansions, the lookup table is a configurable multiple-choice hash table without chains. During lookup, pre-defined hash functions are used one by one, such that if the key is not found, the next hash function is used. If all hash functions are used and the key was still not found then \name returns a miss. Similarly if a collision happens during insertion, the key is re-hashed with the next hash function to map it to another entry in the lookup table. If all hash functions are used and there is still a collision, the last collided object is evicted to make space for the new key.

\begin{figure}[t]
    \centering
	\includegraphics[width=\columnwidth]{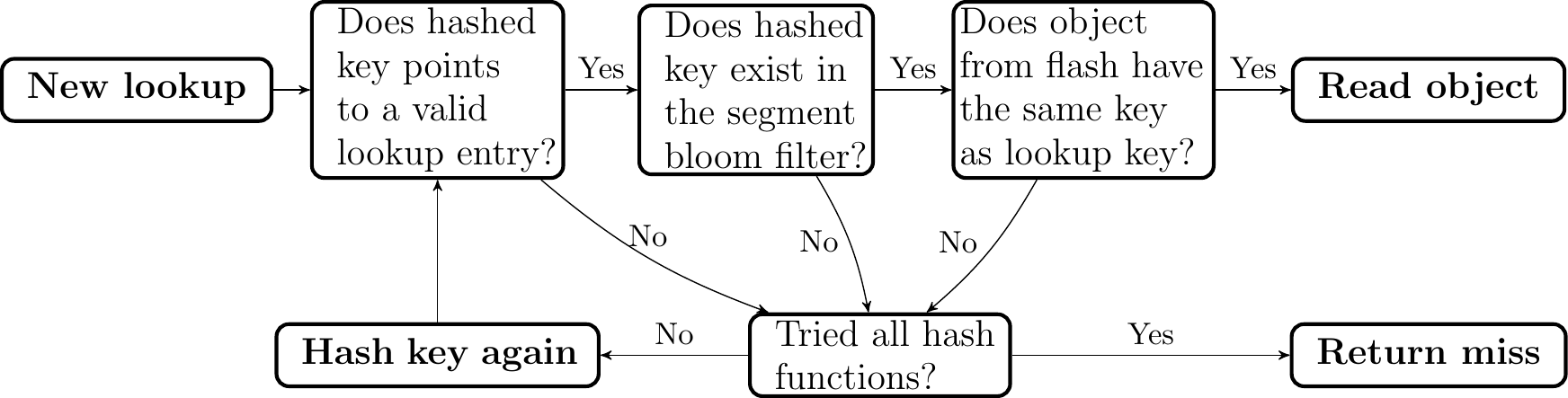}
    \caption{Algorithm for determining if an object exists in flash.}%
    \label{fig:flash_lookup}%
\end{figure}

To reduce the number of excess reads from the flash in case of hash collisions, \name utilizes an in-memory bloom filter for each segment, which indicates whether a key is stored in the segment. We decided to use a bloom filter per segment, rather than a global bloom filter, since each segment is immutable, which eliminates the need to support deletions. We use bloom filters with a false positive rate of 1\%. For the Memcachier trace, this translates to an average of 1.03 accesses to flash for every hit in the flash and an extra memory overhead of 10 bits per item.
Figure~\ref{fig:flash_lookup} summarizes \name's lookup process.

% To find out which segment an object belongs to, the object's key is hashed to the
% corresponding entry in the lookup hashtable, which provides the segment ID.
% Then, \name performs a lookup for the key in the
% segment's bloom filter. If the key is found in the bloom filter, \name reads the
% object from the segment.
% Since the bloom filter may cause a false positive, if the object that was read from flash
% does not have the same key as the object which is being looked up, the key will be hashed
% again and \name will look it up in a different segment.
% Similarly, if the key is not found in the bloom filter, the key is hashed again and \name
% performs another lookup in the lookup hashtable.
% \name will attempt to lookup an object using all the configured hash functions (8 by default) until the object is found. If the object
% is not found after all the attempts, the object does not exist in flash and \name returns a miss.

%\begin{table}[t!]
%\centering
%\footnotesize
%\begin{tabular}{rrrrrr}
%\toprule
%& Segment & Hash & Offset & CLOCK & Ghost \\
%Num Bits & 14 & 3 & 10 & 3 & 1 \\ \bottomrule
%\end{tabular}
%\caption{Flash hashtable entry format.}
%\label{tbl:offset}
%\end{table}

\begin{figure}[t]
    \centering
	\includegraphics[width=\columnwidth]{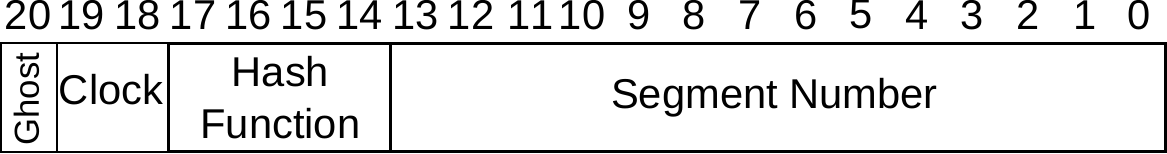}
    \caption{Hashtable entry format for objects stored on flash.}%
    \label{fig:offset}%
\end{figure}

Instead of utilizing a full eviction queue with a linked list of pointers, \name uses the CLOCK
algorithm~\cite{CLOCK}. To evaluate this design choice,
we ran the top 5 applications in the Memcachier trace in a simulation and compared the results between the CLOCK algorithm
and LRU.
The results show that while CLOCK slightly decreases the overall hit rate, the overall effect is negligible.
The results led to us to assign only 2 index bits for the CLOCK algorithm. We describe the CLOCK algorithm in \S\ref{sec:clock}.

\begin{table}[t!]
\centering
\footnotesize
\begin{tabular}{rrrrr}
\toprule
& \multicolumn{4}{c}{Hit Rate} \\
App & CLOCK 1 & CLOCK 2 & CLOCK 3 & LRU \\ \midrule
1 & 70.4\% & 70.57\% & 70.57\% & 70.99\% \\
3 & 99.97\% & 99.97\% & 99.97\% & 99.97\% \\
19 & 99.54\% & 99.55\% & 99.55\% & 99.55\% \\
18 & 98.15\% & 98.15\% & 98.15\% & 98.21\% \\
20 & 99.18\% & 99.22\% & 99.22\% & 99.26\% \\
\bottomrule
\end{tabular}
\caption{Simulated hit rates of CLOCK algorithm approximating LRU with 1, 2, and 3 bits compared to an LRU eviction queue across the top 5 applications in the Memcachier traces.}
\label{tbl:CLOCK}
\end{table}

The hashtable entry format is summarized in Figure~\ref{fig:offset}.
The index contains an extra bit that indicates whether the object is scheduled for deletion from flash (\S\ref{sec:implementation}).

\section{Implementation}
\label{sec:implementation}

\begin{figure}[t]
    \centering
	\includegraphics[width=\columnwidth]{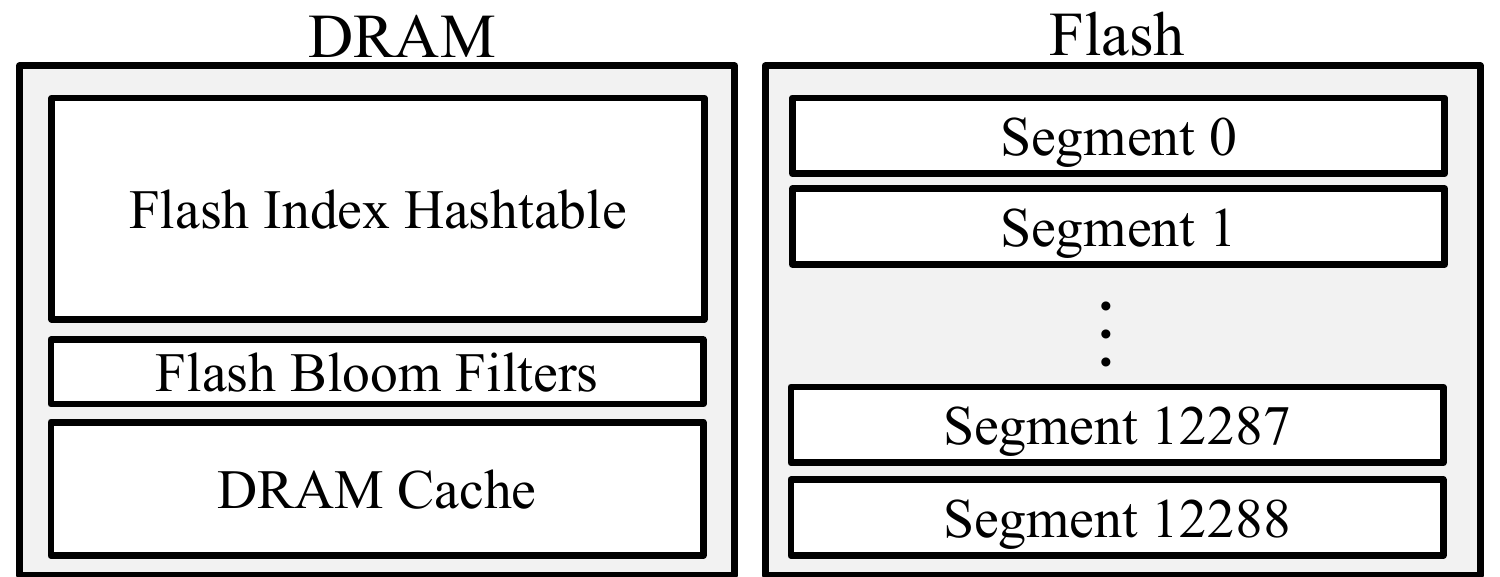}
    \caption{\name's architecture.}%
    \label{fig:architecture}%
\end{figure}

This section presents the implementation details of \name.
We implemented \name in C. Most of the cache functionality was implemented from scratch, except
for the transport, dispatch, request processing, and the hash table for DRAM objects, which are borrowed from Memcached
1.4.15.
\name has four main functions: reads, writes, moving data to flash and eviction.
Figure~\ref{fig:architecture} depicts the high level components of \name's architecture. 

For incoming reads, \name first checks whether the object exists in the hash table for DRAM objects,
which is based on Memcached's hash table.
If not, it checks whether the object exists in flash using a separate hash table for flash objects.
If the object exists either in DRAM or flash, \name returns it, otherwise the request is counted as a miss.
Incoming writes and updates are always written first into DRAM. In the case of updates,
the updated object is written in to DRAM, and the old version is invalidated.
\name maintains free space in the size of a segment (e.g. 512~MB) in DRAM for incoming writes.

\name uses a configurable number of worker threads that process the client requests in parallel.
To maintain enough free space on DRAM, \name utilizes a dedicated cleaner thread.
The cleaner works in the background, and is not part of the critical path for requests.
When the free space on DRAM drops below a segment size, if there are enough objects that meet a
threshold for their flashiness score and the flash write rate limit was not reached, the cleaner will buffer them into a segment and move them into flash.
Objects are moved to flash in an order based on their flashiness score.
When the SSD is full, the cleaner will remove the last segment from flash based on FIFO order.

For eviction, \name maintains a global priority rank for all objects, whether they are stored in DRAM or flash.
Objects are evicted from \name based on this global priority. By default the priority is an approximation of LRU.
If the next object for eviction is in DRAM, \name simply evicts it. If the next object for eviction is in flash,
\name marks it as a \emph{ghost} object, and it will be evicted when its segment is removed from flash.
Note that the movement of data from DRAM into flash is decoupled from eviction. They are
conducted in parallel and use different metrics to rank objects.
Objects that are moved between the flash and DRAM always keep their global
priority ranking. When there are not enough objects in DRAM that meet a threshold for their flashiness score, or the flash write rate reached its limit, the cleaner will evict items from DRAM to maintain sufficient free space.

The rest of the section describes in detail how
\name moves objects into flash, and the implementation of \name's classifier and eviction algorithm.

\subsection{Writing Objects to Flash}
\label{sec:writing}

\name constructs a flash-bound segment in DRAM, by greedily trying to find space for the objects
in the segment one-by-one. The output bits of the pre-determined hash functions provide different
possible insertion points in the segment for each object.
\name first assembles a group of objects that need to be moved to flash based on the their flashiness.
It then tries to insert the objects from this group based on their size. Larger
objects go first, because they require more contiguous space than smaller objects.
In this process, some objects will not have available space in the segment.
\name skips these objects and tries to insert them again next time it creates a new segment.
We evaluate the resulting segment utilization in Section~\ref{sec:utilization}.

\subsection{Classifier Implementation}

\name's flashiness score is computed based on five features for each object, which track information about its past hits.
Since these features depend on information across
multiple object accesses, the features for an object are only generated after an object
has been read at least once. If an object has never been read, its
flashiness score is automatically equal
to zero.

\name periodically trains a separate classifier for each application. For the week-long commercial traces we used, we found that a training
period of one day at the beginning of the trace was sufficient for classifying flashiness for the whole week.

The na\"{\i}ve way to train the classifier would be to update the features at each access to the DRAM.
However, this approach may oversample certain objects, which can create an unbalanced classifier. For example,
if a small set of objects account for 99\% of all accesses, multiple sets of features would be created for these objects,
and the flashiness estimation would be biased towards popular objects.

To tackle this problem, we implemented a sampling technique that generates a single sample for each object,
chosen uniformly over all of its accesses during the training period. Instead of updating the features at each object access, 
with do it only with a probability of 1/n, where $n$ is the number of times the object was read so far.

To illustrate this sampling technique, consider the following example. Suppose an object was written in time $t=0$
and read for the first time at $t=1$. Its features vector will be: $\big[1, 1, 1, 1, 1\big]$
(number of past reads, average time between reads,
time between last two reads, maximum time between subsequent reads, time of first read). Since the number of reads is equal to 1,
the feature vector generated by its first read will be the feature we use for training at a probability of 1. If a second read arrives
at $t=1.5$, then the features after the second read will be: $\big[2, 0.75, 0.5, 1, 1\big]$. \name will keep the second set of features with a probability of 1/2, since the number of reads is equal to 2. This is equal to uniformly sampling the features
from the first or second access.
Each subsequent access will be sampled
at a uniform probability of 1/n, and the probability of prior accesses to be sampled will also be uniform.

After collecting the samples for a day, we measure the number of times each of the objects is hit in the subsequent hour. This number is used as the target function for the training. After these two periods, \name trains the classifier using these training samples and labels.

\subsection{Eviction}
\label{sec:clock}
\name utilizes the CLOCK algorithm to rank objects for eviction.
Each object has two bits in its hash table entry that signify priority.
In order to approximate LRU, when the object is read, its bits are all set to 1.
MFU (Most Frequently Used) is approximated by incrementing the bits by
1 at each read.

Each time it needs to free up space, \name
walks through all the object entries in round-robin order
and decrements their CLOCK entries. It stops decrementing the entries once it reaches
an object that has a CLOCK entry equal to zero, which is the next object for eviction.
If the next object for eviction is in DRAM, it is simply deleted. 
If the object is in flash,
it cannot evict it immediately, since erasing a small amount of data from flash creates write amplification.
Instead, it is marked as \emph{ghost object} which means it is scheduled for eviction once its segment is removed from flash.

\name approximates which objects are at the top of the global eviction rank
(including flash and DRAM). These objects are defined as hot objects.
It maintains a \emph{hot data threshold} to approximate the amount of
hot data in the cache. If the amount of hot data exceeds
the hot data threshold, \name triggers an eviction to reduce it.

The hot data threshold (HDT) is computed by:

$HDT = DRAM + SSD \cdot hot$

Where $DRAM$ is the available capacity of DRAM excluding the lookup table
and free space needed for incoming writes and for buffering data into flash. $SSD$ is
the total size of the SSD, and $hot$ is the percentage of objects on flash that are not ghosts.
By default, $hot$ is set to 70\%, which means that approximately 30\% of the objects
on flash are ghost objects.

Ghost objects can still be accessed after they were marked as ghosts,
since they are not immediately removed from flash. If a ghost object is accessed, we mark it as a hot object (we set
the ghost bit to zero). As a result if the amount of hot data exceeds the hot data threshold,
\name will do a round of decrementing the hash table CLOCK bits,
until it find a sufficient number of flash objects with CLOCK bits of zero, which can be marked as ghosts.
Note that in this case, we do not evict
low ranking objects from DRAM, but only mark flash objects as ghosts.
We do this in order to avoid evicting an object from DRAM after a flash read, which could cause the DRAM to be
underutilized. Therefore, objects from DRAM only get evicted due to incoming writes.

\section{Evaluation}
\label{sec:evaluation}

In this section we evaluate the end-to-end performance of \name compared to existing systems using
the Memcachier traces, measure its
performance using a synthetic microbenchmark, and evaluate the effects of individual trade-offs
we made in its design.

\subsection{End-to-end Performance}

\begin{table}[t!]
\centering
\footnotesize
\begin{tabular}{rrrrrrr}
\toprule
& \multicolumn{2}{c}{\name} & \multicolumn{2}{c}{RIPQ} & \multicolumn{2}{c}{Victim Cache} \\
App & Hit \% & CLWA & Hit \% & CLWA & Hit \% & CLWA \\ \midrule
2  & 98.8\% & 5.8 & 98.5\% & 151.9 & 99.3\% & 4536.3 \\
7  & 98.6\% & 2.8 & 98.8\% & 4.4 & 98.9\% & 21.7 \\
10 & 83.1\% & 0.4 & 83.1\% & 2.9 & 93.3\% & 3.7 \\
20 & 98.1\% & 0.2 & 98.7\% & 12.4 & 99.3\% & 34.0 \\
23 & 96.0\% & 0.8 & 96.0\% & 1.6 & 96.2\% & 1.3 \\
29 & 90.1\% & 0.2 & 91.3\% & 1.8 & 94.4\% & 2.4 \\
31 & 97.3\% & 0.5 & 97.3\% & 1.4 & 97.4\% & 1.0 \\
\bottomrule
\end{tabular}
\caption{Hit rates and CLWA of \name using a threshold of one read future read, RIPQ and victim cache.}
\label{tbl:comparison}
\end{table}

\begin{table}[t!]
\centering
\footnotesize
\begin{tabular}{rrrrrrr}
\toprule
& \multicolumn{2}{c}{\name 1} & \multicolumn{2}{c}{\name 10} & \multicolumn{2}{c}{\name 100} \\
App & Hit \% & CLWA & Hit \% & CLWA & Hit \% & CLWA \\ \midrule
2  & 98.8\% & 5.8 & 99.0\% & 9.2 & 98.9\% & 5.0 \\
7  & 98.6\% & 2.8 & 98.6\% & 2.7 & 95.2\% & 0.0 \\
10 & 83.1\% & 0.4 & 83.1\% & 0.4 & 83.0\% & 0.4 \\
20 & 98.1\% & 0.2 & 98.1\% & 0.2 & 98.1\% & 0.2 \\
23 & 96.0\% & 0.8 & 95.9\% & 0.7 & 95.9\% & 0.7 \\
29 & 90.1\% & 0.2 & 85.5\% & 0.0 & 85.2\% & 0.0 \\
31 & 97.3\% & 0.5 & 97.3\% & 0.5 & 97.3\% & 0.5 \\
\bottomrule
\end{tabular}
\caption{Hit rates and CLWA of \name using a flashiness prediction threshold of 1, 10 and 100 future reads.}
\label{tbl:thresholds}
\end{table}

We compare the end-to-end hit rate and write amplification of \name to RIPQ and the victim cache policy,
by re-running applications from the Memcachier traces against a simulation of the three systems.
Each one of the policies uses the same amount of memory that was allocated in the Memcachier trace,
with a ratio of 1:7 of DRAM and SSD.
We run \name with a threshold of one future read.
In other words, objects
that are predicted to have at least one future read are deemed sufficiently flash-worthy.
Since \name utilizes a separate SVM for each application, we compare the results of individual applications.
To simulate RIPQ with 8 insertion points, and therefore at least 8 different segments on flash, we only run the simulation with
applications that were allocated a sufficient amount of memory by Memcachier.

Table~\ref{tbl:comparison} presents the results of running the simulation against these applications.
The results show that \name achieves significantly lower CLWA than RIPQ and victim cache.
The median CLWA of \name is 0.54, the median of RIPQ is 2.85 and the median of victim cache is 3.67.
Even though \name uses a low threshold for flashiness of one future read,
it still prevents a large number of writes that are not a good fit for SSD from being written to flash.
\name and RIPQ have an almost identical hit rate. Both have a lower hit rate than victim cache, but victim
cache suffers from significantly higher CLWA (and much higher overall WA due to its DLWA).

Table~\ref{tbl:thresholds} compares \name with different flashiness prediction thresholds.
While the results vary from application to application, generally speaking, the higher the threshold
the lower the WA and the lower the hit rate. Note that in some applications, such as in application 2,
this trade off does not hold, since we train the classifier individually on each application,
and each application performs differently.

\begin{table}[t!]
\centering
\footnotesize
\begin{tabular}{rrrrrrr}
\toprule
& \multicolumn{2}{c}{DRAM 1:15} & \multicolumn{2}{c}{DRAM 1:7} & \multicolumn{2}{c}{DRAM 1:3} \\
App & Hit \% & CLWA & Hit \% & CLWA & Hit \% & CLWA \\ \midrule
2  & 99.0\% & 5.1 & 99.0\% & 4.6 & 99.0\% & 2.6 \\
7  & 98.3\% & 3.1 & 98.6\% & 4.1 & 98.8\% & 4.9 \\
10 & 81.4\% & 0.4 & 83.2\% & 0.4 & 92.7\% & 0.8 \\
20 & 97.6\% & 1.2 & 98.4\% & 0.9 & 98.9\% & 2.2 \\
23 & 95.7\% & 0.7 & 96.0\% & 0.8 & 96.2\% & 0.9 \\
29 & 89.0\% & 0.2 & 91.0\% & 0.3 & 94.3\% & 0.4 \\
31 & 97.2\% & 0.5 & 97.3\% & 0.5 & 97.3\% & 0.5 \\
\bottomrule
\end{tabular}
\caption{Hit rates and CLWA of \name using a threshold of 1, with varying ratios of DRAM and SSD. The results use a smaller segment size (2~MB).}
\label{tbl:DRAM}
\end{table}

Table~\ref{tbl:DRAM} depicts the results when we vary the ratio of DRAM and SSD, while keeping the total amount
of memory constant for each application.
The results show that if we reduce the amount of DRAM too much, the hit rate drops. This is due to the fact that when the DRAM is low,
objects do not have enough time to prove themselves as flashy enough to be moved to SSD.
Note that we used a smaller segment size in these runs, in order to display results for a 1:15 ratio of DRAM.

\subsection{Microbenchmarks}

\begin{table}[t!]
\centering
\footnotesize
\setlength\tabcolsep{3.5pt}
\begin{tabularx}{\columnwidth}{Xccc|cc}
\hline
& \multicolumn{3}{c|}{\name} & \multicolumn{2}{c}{Memcached}  \\
& SSD  & DRAM  & \multirow{2}{*}{ Misses} & \multirow{2}{*}{Hits} & \multirow{2}{*}{Misses}\\ 
& Hits & Hits & & & \\ 
\hline
Throughput & \multirow{2}{*}{150,264}& \multirow{2}{*}{270,437} & \multirow{2}{*}{239,191} & \multirow{2}{*}{275,379} & \multirow{2}{*}{286,718}\\
(IOPS)&&&&\\
Latency ($\upmu$s)& 106 & 13.5 & 19 & 13 & 12 \\
\hline
\end{tabularx}
\caption{Throughput and latency of SSD hits, DRAM hits and cache misses for \name and Memcached}
\label{tbl:microbenchmarks}
\end{table}

In the case of both Memcachier and Facebook, Memcached is not CPU bound, but rather memory capacity bound ~\cite{cliffhanger}. Since the Memcachier traces are fairly sparse in terms of their rate of requests, we
ran a set of synthetic microbenchmarks to stress the performance of the system to measure its throughput and latency. 

Our microbenchmarks run on 4-core
3.4 GHz Intel Xeon E3-1230 v5 (with 8 total hardware
threads), 32 GB of DDR4 DRAM at 2133 MHz with a 480~GB Intel~535 Series SSD. All experiments
are compiled and run using the stock kernel, compiler,
and libraries on Debian 8.4 AMD64. 
The microbenchmark requests are based on sequential keys, with the average object size as Memcachier. We disabled the operating system buffer cache to guarantee that SSD reads are routed directly to the SSD drive.
Since the performance of SSD and DRAM is an order of magnitude different, we separately measured SSD and DRAM hits.
Finally, we measured the latency and throughput of Memcached 1.4.15 as a baseline.   

Table~\ref{tbl:microbenchmarks} presents the throughput and latency of the microbenchmark experiment.
The latency and throughput of DRAM hits in \name are very similar to the latency and throughput of Memcached. While the average latency of SSD hits is significantly higher than DRAM, their latencies become similar when deploying over the network (network access times are typically
100~$\upmu$s or more). 
The miss latency of \name is similar to the latency of DRAM hits, because all of \name's lookup indices are stored in DRAM,
and the only case it needs to access flash in a miss is when one of the in-memory bloom filters returns a false positive.  
The write throughput and latency of \name were identical to Memcached, because writes always enter \name's DRAM.

\subsection{Utilization on Flash}
\label{sec:utilization}

When moving data from DRAM to flash, \name tries to allocate space for objects in different possible insertion points
in the flash segment, using pre-defined hash functions. If no space is found for the object, \name skips the object and will try to insert it
next time it moves a segment to flash.

\begin{figure}[t]
    \centering
	\includegraphics[width=\columnwidth]{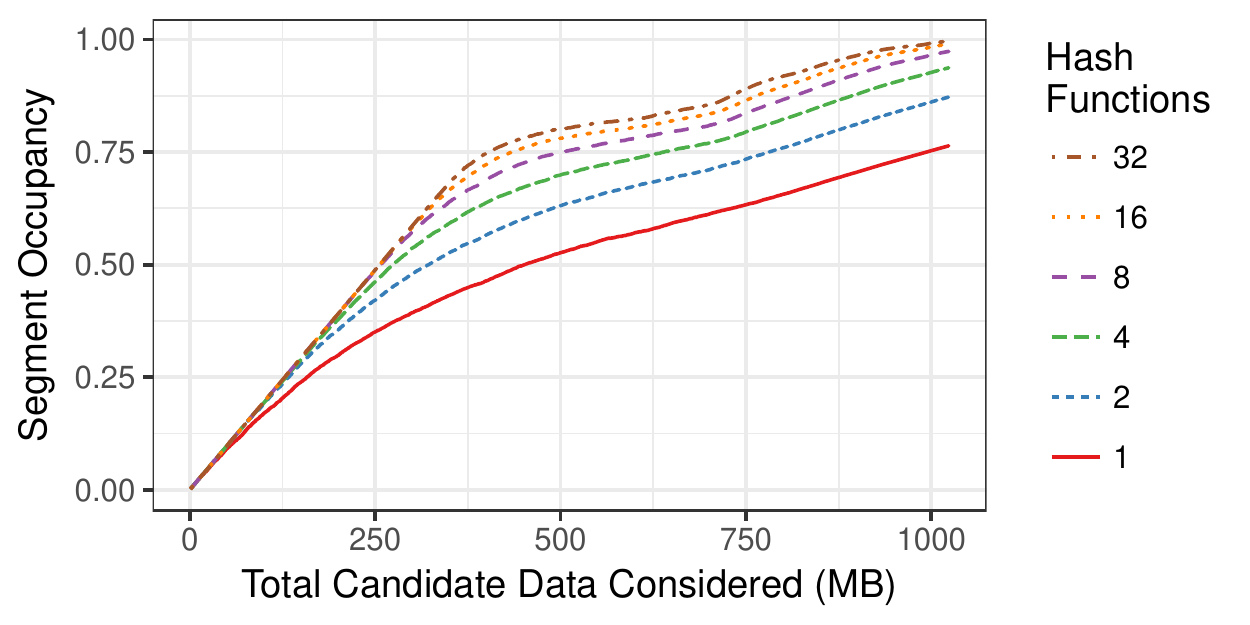}
    \caption{Simulation of the utilization of a 512~MB segment on flash when \name tries to allocate
    space with a varying number of objects from the Memcachier trace. As \name tries to allocate more objects, it achieves
    higher utilization.}%
    \label{fig:segment_util}%
\end{figure}

Figure ~\ref{fig:segment_util} depicts the utilization of \name's flash allocation algorithm. To measure the utilization, we
ran \name's allocation algorithm on the Memcachier trace with different number of hash functions over a segment size of 512~MB. The allocation greedily tries to allocate space to more data and measures the resulting utilization. Note that after the segment reaches about 60\% utilization,
its utilization curve gradient decreases, since when \name tries to allocate objects there is a higher probability of collisions
with other existing objects in the segment. Using 16 hash functions, it takes about 1~GB of objects to reach a 99\% utilization, and on average each object needs to be hashed 8.2 times until it finds an insertion point with enough space.

\section{Related Work}

There are several systems that try to extend the lifetime of flash for the purpose of a key value cache.

\subsection{SSD-based Key Value Caches}

RIPQ~\cite{RIPQ}, Facebook's photo cache, reduces write amplification by buffering data in memory before
writing to flash, and by co-locating similarly prioritized content on flash.
However, RIPQ suffers from more than 5$\times$ higher write amplification than
\name. This is due to two main reasons. First, in RIPQ all content is written to flash. In contrast, by
using DRAM as a filter for objects that are frequently updated or never accessed, \name significantly reduces
the number of writes to flash. Second, since RIPQ constantly co-locates objects with
similar priorities in flash, it frequently rewrites the same objects into flash.
In addition, TAO~\cite{tao}, Facebook's graph data store, uses a limited amount of flash as a victim cache for data stored in DRAM.
Therefore, it suffers from a high rate of writes,
because items which are not frequently
accessed are written into flash.
% For its SSD-based data center cache, Twitter uses Fatcache~\cite{Fatcache}, which is a modified
% version of Memcached that buffers small writes and utilizes FIFO as an eviction policy.
% \name has better write amplification than Fatcache, since not all write requests are written to flash, and higher
% hit rates, because it uses eviction policies like LRU, which provide a higher hit rate than FIFO.

A couple of systems try to support SSD-based caches by modifying the SSD's Flash Translation Layer (FTL).
Duracache~\cite{duracache} tries to extend the life of the SSD cache, by dynamically increasing
the flash device's error correction capabilities. This requires at the minimum modifying the
FTL, and in order to achieve high performance it would require to modify
the ASIC itself. Shen et al~\cite{shen} allow the cache to directly map keys to the device itself,
and remove the overhead of the flash garbage collector. Unlike both of these systems,
\name does not require any changes in the flash device, and addresses the main
cause of write amplification in caches,
which is cache-level write amplification.

In addition, there are several systems that utilize Flash as a block-level cache for
disk storage~\cite{pannier,nitro,flashtier,SDF,flashcache,icache}.
In particular, Pannier~\cite{pannier} and Nitro~\cite{nitro} are block-level caches that reduce
write amplification by caching objects that are read frequently and updated
infrequently. However, unlike \name, they do not utilize DRAM as a filter for SSD to further
reduce write amplification.

Cheng et al~\cite{belady} present an offline analysis of the trade-off between write amplification and
eviction policies in block-level caches. They generalize Belady's MIN algorithm to flash-based caches, and demonstrate that LRU-based eviction is still far from their proposed optimal oracle eviction policy.
However, they do not provide an online algorithm and an implementation that
reduces write amplification of SSD-based caches.

\subsection{SSD-based Key Value Stores}

There are many key-value storage systems designed for SSD. These are typically not suitable for
the cache use case, since they incur high levels of write amplification.

For example, LevelDB~\cite{leveldb} and RocksDB~\cite{rocksdb} are key-value stores
based on Bigtable~\cite{bigtable} that are frequently deployed on SSD devices. Both of these stores
incur a write amplification of more than 3$\times$ (and even as high as 10$\times$ or more)~\cite{MSLD,wisckey,LSM-trie},
because they use Log-structure Merge-trees (LSM), and incur an
additional write each time an object moves to a new level.

WiscKey~\cite{wisckey} improves the performance and write amplification of LevelDB by only
sorting the keys, without sorting the values in the log. Even though it significantly reduces the
write amplification of LevelDB, it still suffers from up to 4-5$\times$ write amplification
when the workload contains small values. Similarly, LSM-trie~\cite{LSM-trie} also improves
the performance and write amplification of LevelDB by leveraging a trie to make compaction more efficient. 
However, it still suffers from a write amplification of up to 5$\times$, and requires
two accesses to flash for each read from flash.

SILT~\cite{SILT} is a flash database that minimizes the index
stored in memory, by utilizing space efficient indexing techniques, like cuckoo hashing and
entropy-coded tries. We were inspired by some of these techniques in the design of \name's flash index.
However, unlike \name, SILT is not optimized for write amplification: in order
to compress the in-memory index, each object is written more than twice on flash, and up to 20 times under
certain workloads~\cite{LSM-trie}. In addition,
unlike \name, SILT assumes fixed sized objects.
\section{Conclusions}

SSD faces unique challenges to its adoption as a key-value
cache, since the small object sizes and the frequent rate of evictions and updates creates excessive writes and erasures
on flash storage.
We presented \name, the first key-value cache that uses DRAM as a filter for objects
that are not ideal for SSD. Our main insight was that many of the objects written to key-value caches are never read and frequently updated,
and can be filtered dynamically.
\name profiles objects using light-weight machine learning, and dynamically learns and predicts
which objects are the best fit for flash storage.
To efficiently utilize the DRAM both as a filter and as a cache, we designed
a novel in-memory index that supports variable objects with an overhead of less than 4 bytes
per object.
We implemented \name, and showed that it reduces write amplification to a median of 0.5 compared to existing systems,
which suffer from 5$\times$ more write amplification, and much shorter SSD lifetimes.

\footnotesize \bibliographystyle{abbrv}
\bibliography{bib}

\end{document}